\documentclass[pra, amsmath, amssymb, aps, floatfix, showkeys, twocolumn, superscriptaddress]{revtex4-2}

\usepackage{graphicx}
\usepackage{dcolumn}
\usepackage{bm}
\usepackage[colorlinks, linkcolor = magenta, urlcolor = blue, citecolor=blue]{hyperref}
\usepackage{braket}
\usepackage{physics}
\usepackage{amsmath}
\usepackage{array}
\usepackage{float}
\usepackage{multirow} 
\usepackage{caption}
\begin{document}
\title{Realizing Negative Quantum States with the IBM Quantum Hardware}
\author{Jai Lalita}
\email{jai.1@iitj.ac.in}
\affiliation{Indian Institute of Technology, Jodhpur-342030, India}
\author{Pavithran Iyer}
\email{pavithransridhar@gmail.com}
\affiliation{Institute for Quantum Computing, University of Waterloo, Canada}
\author{Subhashish Banerjee}
\email{subhashish@iitj.ac.in}
\affiliation{Indian Institute of Technology, Jodhpur-342030, India}

\newcommand{\pavi}[1]{\textcolor{red}{#1}}

\date{\today}
             
\begin{abstract}
This study explores robust entangled states described using the framework of discrete Wigner functions. Notably, these states are known to outperform the Bell state in measures of entanglement in the presence of non-Markovian noise. Our study focuses on methods for preparing these states using quantum circuits that can be implemented on superconducting hardware and testing the efficacy of these methods on IBM’s quantum device. We present quantum circuits for state preparation and validate them through tomographic reconstruction on the IBM \emph{ibm\_brisbane} device. We propose a teleportation scheme that leverages these entangled states as a resource. We believe that these entangled states have the potential to be used in place of the traditional Bell state in scenarios where non-Markovian errors are prevalent.
\end{abstract}


\maketitle

\section{Introduction}
Entanglement is a unique ingredient in a quantum computing scheme that is pivotal to harness the full power of quantum mechanics. In addition to being a resource for speeding up computations, entanglement within quantum systems is also responsible for the robust storage of quantum information in a quantum memory~\cite{Quantumerrorcorrection2013}. Since the advent of quantum computing, considerable efforts have been directed towards attaining a profound understanding of the entanglement properties inherent in quantum systems. Classical and quantum systems can be described using a quasi-probability distribution of their configurations in the phase space. In contrast to classical systems, the quasi-probability associated with a quantum mechanical system, known as the Wigner function \cite{W32}, can also assume negative values. In fact, the presence of negativity in the values of the Wigner function is widely used as a test of non-classicality in a system~\cite{AZ04}. Although the Wigner function is conventionally applied in the context of continuous variable systems, there exists a less prevalent but potent theory concerning the discrete Wigner function (DWFs)~\cite{wootters2004picturing, gibbons2004discrete}. Analogously to the continuous variable scenario, the states of a system whose discrete Wigner function manifests negative values within the discrete phase space can be represented as eigenstates of the negative eigenvalues of phase-space point operators~\cite{gibbons2004discrete,van2011noise, casaccino2008extrema}.

Noise is an imminent threat to quantum memories. The impact of noise on quantum information can be assessed by examining the open quantum dynamics of a system, which is characterized by a master equation. A general solution of the master equation that accommodates a broad range of real-world noise is articulated by the space of non-Markovian maps~\cite{banerjee2018open, breuer2002theory, SM23, TB23}. Quantum error correction, intended to ensure the resilience of quantum information encoded within memory to noise, is frequently presumed to be effective under highly simplified assumptions about the physical noise process, such as depolarizing noise. Consequently, it remains largely uncertain yet crucial to determine whether quantum error correction and, more fundamentally, entanglement can be effectively utilized for defense against non-Markovian errors. 

In a recent study, quantum states that exhibit negativity of the discrete Wigner function have been identified to be robust to a wide variety of noise in quantum systems captured by non-Markovian errors~\cite{lalita2023harnessing, Lalita_2024ProtectingQC}. These two-qubit negative quantum states are recognized as optimal candidates for universal quantum teleportation using weak measurements~\cite{Lalita_2024ProtectingQC} within non-Markovian noise environments. Consequently, the physical realization of these two-qubit negative quantum states is imperative, as they hold the potential to augment quantum information protocols by providing resilient resources.

In this work, we present methods to generate these states using operations native to superconducting hardware, specifically single-qubit gates and the $CZ$ gate. Through quantum state tomography, we demonstrate that these states can be prepared with high fidelity on IBM’s quantum hardware under realistic noise conditions. We employed various approaches to assess the noise resilience of the negative quantum states and benchmarked them against the standard Bell states. These approaches include fidelity variation analysis, phase sensitivity under $SU(2)$ rotations using quantum Fisher information (QFI), violations of Bell-CHSH inequality, and quantum information measures such as concurrence and teleportation fidelity. Our results highlight the role of negative quantum states as robust entanglement resources for a range of quantum computing and communication applications. Furthermore, we propose a teleportation circuit that utilizes one of the two-qubit negative quantum states as its entanglement resource.

This paper is organized as follows. Section~\ref{sec:preli} describes the negative quantum states within the framework of discrete Wigner functions and discusses quantum state tomography. In Sec.~\ref{sec:methods}, we present methods for preparing these states on IBM superconducting hardware along with ideal and mitigated state tomography techniques. Section~\ref{sec:results} tests negative quantum states' fidelity, maximal mean QFI, and Bell-CHSH inequality violation under non-Markovian noise. Also, a teleportation circuit using one of the negative quantum states is provided in this section. This is followed by conclusions in Sec.~\ref{sec:conclusion}.

\section{Preliminaries}\label{sec:preli}
\subsection{Negative quantum states} \label{subsec:neg_states}
The framework presented in~\cite{wootters2004picturing, gibbons2004discrete, lalita2023harnessing, Lalita_2024ProtectingQC} is utilized for the
evaluation of discrete Wigner functions (DWFs) for quantum systems with a Hilbert space dimension of $d = p^n$, where $p$ is a prime number and $n$ is any integer.
We assign a state $\ket{\phi_{i, j}}$ ($\ket{\phi_{i, j}}$'s are the bases vectors of mutually unbiased bases (MUBs)) to each line $\lambda_{i, j}$, $\textit{viz.}$, the $j^{th}$ line of the $i^{th}$ striation in the discrete phase space~\cite{wootters1989MUB, Lawrence2002MUB, bandyopadhyay2002MUB, durt2010mutually}. Once these associations are made, we define a phase space point operator ${A}_{\alpha}$ at every point of the discrete phase space $\alpha(q, p)$ as given in Eq.~\eqref{A_formula}~\cite{wootters1987wigner}. Each such assignment is called a quantum net~\cite{gibbons2004discrete}. For a $d$-dimensional quantum system, there are $d^{d+1}$ possible quantum nets. Thus, $d^{d+1}$ possible definitions of DWFs corresponding to the same density matrix. The Wigner function is a quasi-probability distribution function that allows the phase space representation of a quantum system. The DWFs at phase space point $\alpha(q, p)$ should be such that the sum of $W_{\alpha}$'s over-the-line $\lambda_{i, j}$ gives the probability of finding the quantum system in the state ${P}_{i, j} = \ket{\phi_{i, j}}\bra{\phi_{i, j}}$, i.e.,$
\sum_{\alpha \in \lambda_{i,j}} W_{\alpha} = {\rm Tr}({P}_{i, j}{\rho}). $
Further, the resultant DWFs can be found as $ W_{\alpha} = \frac{1}{d} {\rm Tr}({A}_{\alpha}{\rho} )$~\cite{gibbons2004discrete},
where
\begin{equation}
    \begin{aligned}
      {A}_{\alpha} = \sum_{\lambda_{i,j} \ni \alpha} {P}_{i,j} - {I}.
    \end{aligned}\label{A_formula}
\end{equation}
Here, ${P}_{i,j}$'s are the projectors associated with the lines $\lambda_{i,j}$'s.
Using DWFs, negative quantum states were introduced in~\cite{lalita2023harnessing}. These are the normalized eigenvectors corresponding to negative eigenvalues of $A_{\alpha}$~\cite{van2011noise, casaccino2008extrema, lalita2023harnessing}. The eigenstate associated with the most negative eigenvalue of $A_{\alpha}$ is called the first negative quantum state, denoted by $\ket{NS_1}$. Similarly, the second and third negative quantum states are denoted by $\ket{NS_2}$ and $\ket{NS_3}$, respectively, which correspond to the normalized eigenvectors of the second and third negative eigenvalues of the operator $A_{\alpha}$. This pattern continues for subsequent states of other negative eigenvalues.
\subsection{Quantum State Tomography}
Tomography is a technique used to construct an image of a hidden object by analyzing several observable projections~\cite{Quantum-State-Tomography1995}. Due to the inherent nature of quantum physics, it is not feasible to directly observe physical objects in their actual state. Instead, we see only the various aspects of the physical objects, like the wave or the particle aspects, which depend on the particular type of measurement. However, it is still possible to determine the quantum state by doing numerous experiments on identically prepared systems and building up good statistics on the outcomes. Suppose that the collection of experiments is completely informative. In that case, it is possible to reconstruct the density matrix of the quantum system, known as quantum state tomography. An infinite number of perfect measurements would be required to determine the state entirely. 
\section{Methods}\label{sec:methods}
\subsection{Unitary transformations to realize negative quantum states}
In this work, we particularly implement two-qubit negative quantum states elaborated in~\cite{lalita2023harnessing, Lalita_2024ProtectingQC} on IBM's quantum hardware. In the case of two-qubit quantum systems, the number of possible $A_{\alpha}$'s and probable DWFs is $4^{4+1}$. Among all, 320 $A_{\alpha}$'s exhibit the spectrum ($-0.5000$, $-0.5000$, $0.1339$, $1.866$), another 320 have the spectrum ($-0.8661$, $-0.5000$, $0.8661$, $1.5000$), and the remaining 384 have the spectrum ($-0.8968$, $-0.1420$, $0.2787$, $1.7601$) as described in~\cite{casaccino2008extrema}. By exploring various combinations of MUB vectors and striations, we identify three distinct $A_{\alpha}$'s, each having one of the above possible spectra. The operator $A_{\alpha}$ at the phase space point $\alpha(1, 1)$ with spectrum ($-0.8968$, $-0.1420$, $0.2787$, $1.7601$) is given by
\begin{equation}
   \begin{aligned}
    A_{(1, 1)} = \left(
\begin{array}{cccc}
 0 & -\frac{1}{2}-\frac{i}{2} & \frac{1}{2}-\frac{i}{2} & -\frac{1}{2} \\
 -\frac{1}{2}+\frac{i}{2} & 0 & \frac{i}{2} & 0 \\
 \frac{1}{2}+\frac{i}{2} & -\frac{i}{2} & 1 & 0 \\
 -\frac{1}{2} & 0 & 0 & 0 \\
\end{array}
\right).
      \end{aligned}
      \label{A_NS1}
\end{equation}
The $\ket{NS_1}$ state is the normalized eigenvector corresponding to the above phase space point operator's most negative eigenvalue $-0.8968$. Similarly, the $\ket{NS_2}$, $\ket{NS_3}$, and $\ket{NS_3^{\prime}}$ states can be derived from the phase space point operators' eigenvalue spectrum. The spectrum ($-0.5000$, $-0.5000$, $0.1339$, $1.866$) possess eigenvalue $-0.5$ with multiplicity $2$. Thus, the eigenvalue $-0.5$ has multiple linearly independent standard eigenvectors. The approximate explicit expressions of the two-qubit negative quantum states are
\begin{eqnarray}
    \begin{aligned}
        \ket{NS_1} &= \left(~a~~b~~c~~d~\right)^T;
        \ket{NS_2} = \left(~p~~q~~r~~s~\right)^T; \nonumber \\
        \ket{NS_3} &= \left(~-l~~m~~n~~l~\right)^T,
        \ket{NS_3^{\prime}} = \left(~-x~~y~~z~~x~\right)^T, \nonumber \\
        \text{and}~~\ket{NS_3^{\prime\prime}} &= \left(~0~~i k~~k~~0~\right)^T.                            
    \end{aligned}
    \label{negative_quantum_states}
\end{eqnarray}
Here, $a = -0.743$, $b = -0.357(1 - i)$, $c = 0.102(1 + i)$, $d = -0.414$, $p = 0.788$, $q = -0.288(1 - i)$, $r = -0.288(1 + i)$, $s = -0.211$, $l = 0.0508$, $m = 0.631 - 0.228i$, and $n = -0.279 - 0.682i$, and $x = 0.575$, $y = -0.346 + 0.310i$, $z = -0.265 - 0.229i$ are the numerically rounded values and $k = \frac{1}{\sqrt{2}}$.

To determine the unitary transformations of the two-qubit negative quantum states, the Gram-Schmidt procedure is employed~\cite{nielsen2010quantum}. A comprehensive description of this procedure is as follows. \textit{Step 1}: Consider any two-qubit negative quantum states, i.e., either of the $\ket{NS_{1}}$, $\ket{NS_2}$, $\ket{NS_3}$, and $\ket{NS_3^{\prime}}$ as $\ket{V_1}$. Let $\ket{V_1}$ = $\ket{NS_1}$. \textit{Step 2}: To find the other three orthonormal vectors ($\ket{V_2}$, $\ket{V_3}$, $\ket{V_4}$) of $\ket{V_1}$, we take any three linearly independent vectors of $\ket{V_1}$. We pick the standard computational basis vectors $\ket{e_{1}} = \ket{00}, \ket{e_{2}} = \ket{01}, \ket{e} = \ket{10}$ as linearly independent vectors of $\ket{V_1}$. Considering $\ket{O_2} = \ket{e_{1}}$, $\ket{O_3} = \ket{e_{2}}$, $\ket{O_4} = \ket{e_{3}}$, the orthonormal vectors $\ket{V_2}$, $\ket{V_3}$, and $\ket{V_4}$ can be calculated using the Gram-Schmidt decomposition as
\begin{equation}
    \ket{V_{K+1}} = \frac{\ket{O_{K+1}} - \sum_{i=1}^{K}\bra{V_i}\ket{O_{K+1}}\ket{V_i}}{||\ket{O_{K+1}} - \sum_{i=1}^{K}\bra{V_i}\ket{O_{K+1}}\ket{V_i}||}.
\end{equation}
\textit{Step 3}: Now, we have four orthonormal vectors $\ket{V_1}$, $\ket{V_2}$, $\ket{V_3}$, and $\ket{V_4}$, which can span the two-qubit system's Hilbert space.
\textit{Step 4}: Finally, the unitary transformation $U$ from the computational basis set $\{\ket{e_i}\}$ to the orthonormal set $\{\ket{V_i}\}$ is given by
     $U = \sum_{i = 1}^{4} \ket{V_i}\bra{e_i}$, which takes the vector $\ket{00}$ to the state $\ket{NS_1}$.
Similarly, the unitary transformations of the other two-qubit negative quantum states can be calculated. The unitary transformation matrices for all the above-mentioned two-qubit negative quantum states are presented in Appendix~\ref{Unitary transformations}. Now, we discuss the quantum circuits corresponding to the unitary transformations of the two-qubit negative quantum states.
\subsection{Circuits for preparing entangled states} \label{subsec:circs}
After obtaining the unitary transformations of negative quantum states, the final optimized circuits for $\ket{NS_1}$, $\ket{NS_2}$, $\ket{NS_3}$, and $\ket{NS_3^{\prime}}$ states using $H$, $R_x$, $R_z$, and $CZ$ gates are constructed and verified using Qiskit~\cite{qiskit2024}. Their corresponding quantum circuits are detailed in Appendix~\ref{Unitary transformations} (Fig.~\ref{NS1_NS2_NS3__NS3_prime_circuit}). Additionally, we can observe that the $\ket{NS_3^{\prime\prime}}$ state has a relative phase of $\pi/4$ with respect to the Bell $\ket{\psi^{+}}$ state, see Fig.~\ref{Teleportation_NS3_double_prime}. Moreover, the circuit depth for $\ket{NS_1}$, $\ket{NS_2}$, $\ket{NS_3}$ and $\ket{NS_3^{\prime}}$ is $13$, while for $\ket{NS3^{\prime\prime}}$ it is just $4$. The loss function for all the quantum circuits is also calculated using $\delta = \sqrt{1 - \frac{|{\rm Tr}(U_f^{\dag}U_t)|^2}{d^2}}$~\cite{Younis2020QFASTQS}, where $U_f$ is the operation implemented by the encoded circuit, $U_t$ is the target input, and $d$ is the dimension of the unitary matrix under consideration. On implementing the circuits given in Fig.~\ref{NS1_NS2_NS3__NS3_prime_circuit} (in Appendix~\ref{Unitary transformations}) on Qiskit's \emph{AerSimulator}, the loss function for all the two-qubit negative quantum states comes out to be $\sim 10^{-5}$. The Schmidt rank of the two-qubit negative states and output states of their corresponding quantum circuits is the same, i.e., $2$. In addition, we perform the tomographic reconstruction of the given quantum circuits on both the simulator and the IBM quantum computer to validate them.
 
\begin{figure}[b]
    \centering
    \includegraphics[width=1\columnwidth]{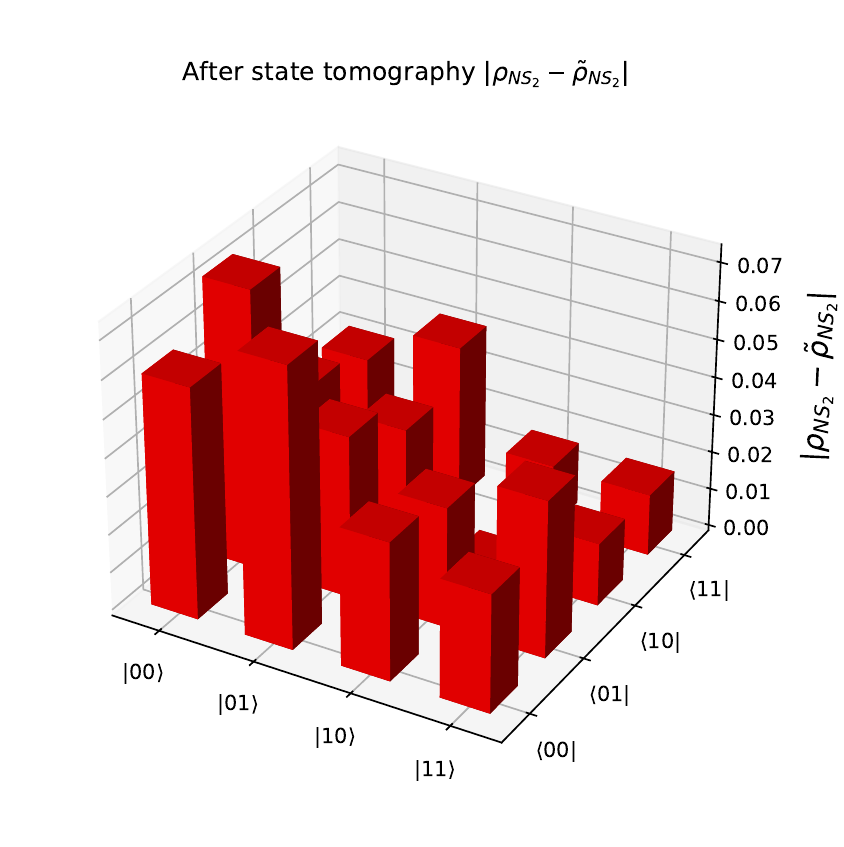}
    \caption{The city plot for the $\ket{NS_2}$ state displays the absolute difference between the components of the original $\rho_{NS_2}$ and the $\tilde\rho_{NS_2}$ obtained after performing a state tomography experiment on the real IBM quantum computer $\it{ibm\_brisbane}$ for $8192$ times.}
    \label{city_plot_NS2_tomo}
\end{figure}

Using Qiskit~\cite{qiskit2024}, we perform the state tomography of two-qubit negative quantum states based on identifying the maximum-likelihood state that aligns with the available data~\cite{Maximum-Likelihood_2012}. The tomography of two-qubit negative quantum states is experimentally implemented by executing their respective quantum circuits for $8192$ times on both the $\textit{ibm\_brisbane}$ quantum computer and the IBM $\textit{AerSimulator}$~\cite{qiskit2024, ibm_brisbane}, resulting in the $\tilde\rho$ state. Figure~\ref{city_plot_NS2_tomo} shows the city plot illustrating the absolute difference between the components of the original state $\rho_{NS_2}$ and the reconstructed state $\tilde\rho_{NS_2}$. As we can observe, the individual elements of the density matrix obtained by state tomography are closer to the actual $\ketbra{NS_2}{NS_2}$ density matrix with a maximum difference of approximately $0.07$. Similarly, we can perform and study the quantum state tomography for other negative quantum states. Further, to quantify the closeness between the state obtained by tomography and the actual negative quantum states, we study their fidelity,~\textit{viz}, given as~\cite{jozsa1994fidelity, nielsen2010quantum} 
\begin{equation}
    F(\rho_{NS_2}, \tilde\rho_{NS_2}) \equiv {\rm Tr}\sqrt{\sqrt{\rho_{NS_2}}\tilde\rho_{NS_2}\sqrt{\rho_{NS_2}}},
    \label{Fidelity_formula}
\end{equation}
Fidelity estimates the likelihood that one state will successfully undergo a test to be recognized as the other, providing a way to validate the quantum circuits of the negative quantum states. The fidelity values for all the negative quantum states together with the Bell state are provided in TABLE I. After tomographic reconstruction, we observe that it is around $0.87-0.91$ on the actual quantum computer and around $0.91-0.93$ on the simulator for the negative quantum states. 
Interestingly, the circuit depth for the negative quantum states is comparatively higher than the Bell state, yet the fidelity values are approximately the same.  
\begin{figure}[b]
    \centering
    \includegraphics[width=1\columnwidth]{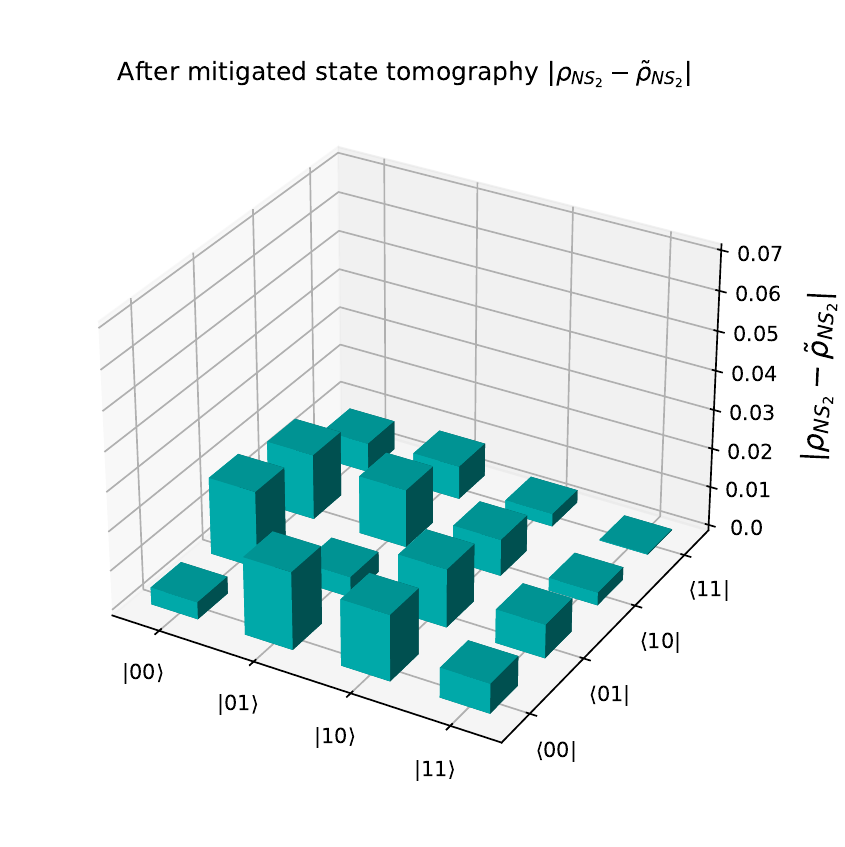}
    \caption{The city plot for the $\ket{NS_2}$ state displays the absolute difference between the components of the original $\rho_{NS_2}$ and the $\tilde\rho_{NS_2}$ obtained after performing a mitigated state tomography experiment on the real IBM quantum computer $\it{ibm\_brisbane}$ for $8192$ times.}
    \label{city_plot_NS2_miti}
\end{figure}

\begin{figure}
    \includegraphics[width=1\columnwidth]{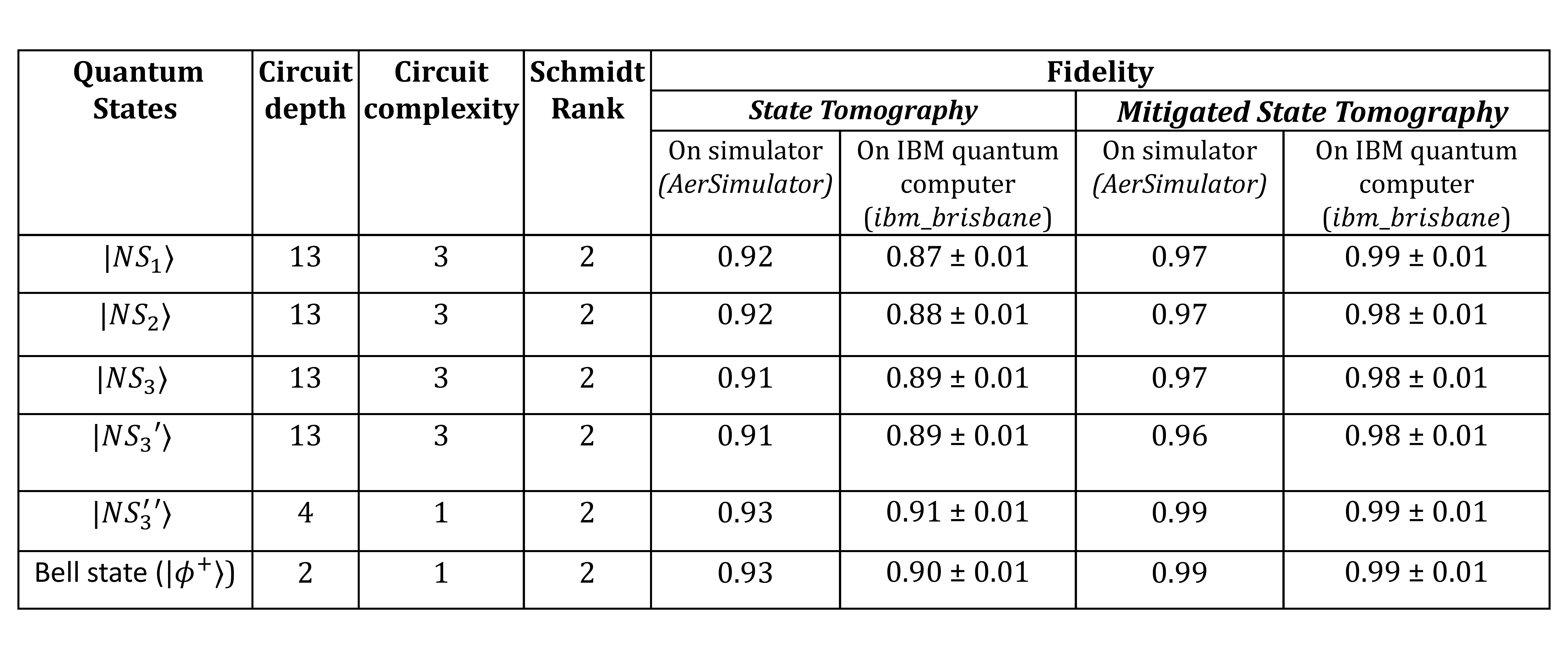}
    \captionsetup{labelformat=empty}
    \caption{TABLE I: Comparison of circuit depth, complexity, Schmidt rank, and fidelity after performing state tomography and mitigated state tomography on IBM $\textit{AerSimulator}$ and real quantum computer $\textit{ibm\_brisbane}$ of the negative quantum states and the Bell state ($\ket{\phi^{+}}$).}
    \label{New_table}
 \end{figure}

Furthermore, readout errors can impact quantum computation when measuring qubits in a quantum device~\cite{ReadoutError2024}. By understanding these errors, we can create a readout error mitigator to improve the accuracy of output distributions and the precision of measurable expectations. We conduct a batched experiment to characterize readout errors and then perform the state tomography. This is known as mitigated state tomography. We carry out the mitigated state tomography for the negative quantum states and the Bell state on the $\textit{ibm\_brisbane}$ quantum computer and $\textit{AerSimulator}$ with 8192 shots. The absolute difference between the components of the original $\rho_{NS_2}$ and the $\tilde\rho_{NS_2}$ obtained after performing mitigated state tomography experiments on the real IBM quantum computer $\it{ibm\_brisbane}$ is depicted in Fig.~\ref{city_plot_NS2_miti}. We observe that the individual matrix elements obtained by mitigated state tomography are closer to the actual values than those obtained by the state tomography. The other negative quantum states also show similar behavior. This improvement is quantified using fidelity and shown in TABLE I. On the simulator, the fidelity is around $0.96-0.99$, while on the actual quantum computer, it is around $0.98-0.99$. Again, after mitigated state tomography, we observe that the fidelity of the Bell state is similar to that of the negative quantum states, though the circuit depth for the negative quantum states is comparatively higher. 
\section{Results}\label{sec:results}
\subsection{Error models and fidelity estimation}
We extend our analysis to assess the robustness of these states under depolarizing noise, as well as under non-Markovian random telegraph noise (RTN) and amplitude damping (AD) noise.

\textit{Depolarizing noise.---} 
Depolarizing noise is a prevalent type of quantum noise in quantum computing. For single qubit, the depolarizing noise channel, $\mathcal{E}(\rho)$, with error probability $p$ can be written as~\cite{nielsen2010quantum},
\begin{equation}
    \mathcal{E(\rho)} = (1 - p)\rho + \frac{p}{3}( X \rho X + Y \rho Y + Z \rho Z) = \frac{p}{2}I + (1 - p)\rho,
\end{equation}
where $X$, $Y$, and $Z$ are Pauli operators and $\rho$ is the density matrix of a qubit. This channel depolarizes the qubit with probability $p$ and leaves the qubit intact with probability $(1 - p)$. 
% The Kraus operators of this channel are
% \begin{equation}
%     K_{0} = \sqrt{1 - p}I, K_{1} = \sqrt{\frac{p}{3}}X, K_{2} = \sqrt{\frac{p}{3}}Y, K_{3} = \sqrt{\frac{p}{3}}Z.
% \end{equation}
We apply the depolarizing noise after constructing the Bell and two-qubit negative quantum states. To this effect, a random variable $r$ uniformly distributed in the interval $[0,1]$ is generated. Whenever the condition $r < \frac{p}{3}$ is satisfied, a Pauli gate ($X, Y$, or $Z$) is randomly selected and applied to both qubits of the state to simulate the occurrence of an error. The Shor's error correction scheme~\cite{Shor_correction} is subsequently implemented to mitigate these induced errors on both the qubits of the two-qubit state, whereas, during error correction, the depolarizing error is randomly applied on one of the qubits, including the ancillary qubits, restoring the integrity of the quantum information.

To observe the effect of depolarizing noise on the generation of the two-qubit negative quantum states and the Bell state, variation of (1 - $F$) ($F$ is the fidelity between the original state and the state gone through the above process) of the $\ket{NS_1}$, $\ket{NS_2}$, $\ket{NS_3}$, and the $\ket{\phi^{+}}$ Bell state with error probability $p$ is calculated. 
\begin{figure}[h]
    \centering
    \includegraphics[width=1\columnwidth]{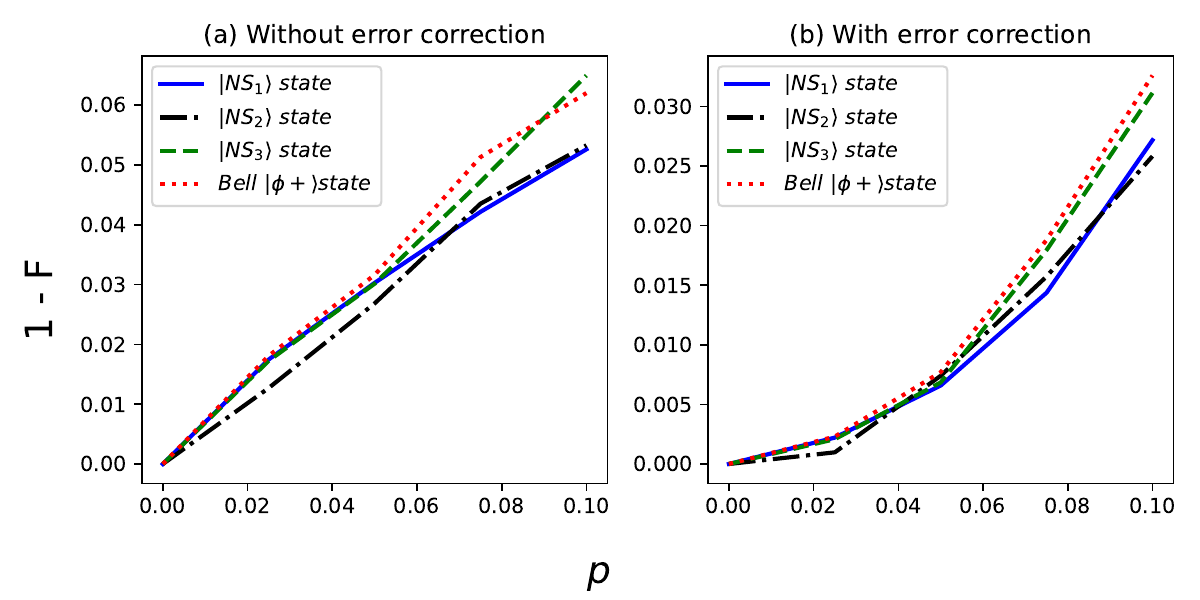}
    \caption{Variation of ($1 - F$) for $\ket{NS_1}$, $\ket{NS_2}$, $\ket{NS_3}$, and $\ket{\phi^{+}}$ Bell state with depolarizing error probability $p$ (a) without any error correction and (b) after implementing Shor’s error correction.}
    \label{DP_NS1_NS2_NS3}
\end{figure}
Figure~\ref{DP_NS1_NS2_NS3}(a) depicts that the fidelity of all the states decreases sharply without performing any error correction. Conversely, Fig.~\ref{DP_NS1_NS2_NS3}(b) illustrates that after implementing Shor's error correction, all the states remain unaffected by depolarizing noise up to an error probability $0.012$ while for the $\ket{NS_2}$ state this value is $0.025$. It shows that the $\ket{NS_2}$ state is robust among all. Furthermore, the behavior of $\ket{NS_3}$ and the Bell state is analogous, but the $\ket{NS_1}$ state dominates over both with increased $p$.

\textit{Non-Markovian noise.---} 
Non-Markovian noise refers to noise in quantum systems where the system retains the memory of its past states, unlike Markovian noise, which assumes memory-less interactions~\cite{breuer2002theory}. In a non-Markovian environment, information lost to the surroundings can flow back into the system, leading to correlations between the current and previous states. Understanding non-Markovian noise is essential for accurately simulating and mitigating errors in quantum computing. Here, we examine the effect of non-Markovian random telegraph noise (RTN) and amplitude damping (AD) noise on the generation of two-qubit negative quantum states and the $\ket{\phi^{+}}$ Bell state. The evolution of a two-qubit system having local interactions with the non-Markovian noisy channels is given by ${\rho}_{AB}(t) = \sum_{i = 0}^{1}\sum_{j = 0}^{1} ({K}_i \otimes {K}_j) {\rho}_{AB}(0) ({K}_i \otimes {K}_j)^{\dag}$, where ${K}_i$'s, and ${K}_j$'s are the Kraus operators of the non-Markovian RTN and AD noise as stated below,
\begin{eqnarray}
      {K_0^{RTN}} = \sqrt{\frac{1 + \Lambda(t)}{2}}{I},
      {K_1^{RTN}} = \sqrt{\frac{1 - \Lambda(t)}{2}}{\sigma_z},
\label{NMRTN_Kraus_operators}      
\end{eqnarray}
\begin{equation}
    {K_0^{AD}} = \begin{pmatrix}
     1 & 0\\
     0 & \sqrt{1 - \lambda(t)}
    \end{pmatrix},
    {K_1^{AD}} = \begin{pmatrix}
    0 & \sqrt{\lambda(t)}\\
    0 & 0
\end{pmatrix}.
\label{NMAD_Kraus_operators}
\end{equation}
Here, $\Lambda(t)$ and $\lambda(t)$ are given by,
\begin{equation}
      \Lambda(t) = e^{-\gamma^{RTN} t}\left[ \cos\left(\zeta \;\gamma^{RTN} t\right) + \frac{\sin\left(\zeta \;\gamma^{RTN} t\right)}{\zeta}\right],
\end{equation}
\begin{equation}
    \lambda(t) = 1 - e^{-gt}\left(\frac{g}{l} \sinh{\frac{lt}{2}} + \cosh{\frac{lt}{2}}\right)^2,
\end{equation}
where $\zeta = \sqrt{\left(\frac{2b}{\gamma^{RTN}}\right)^2 - 1}$, and $b$, $\gamma^{RTN}$ quantifies the RTN's system–environment coupling strength and fluctuation rate. The dynamics is Markovian if {$(4 b \tau)^2 < 1$ and non-Markovian if $(4 b \tau)^2 > 1$ (here, $\tau = \frac{1}{2\gamma^{RTN}}$ as discussed in~\cite{kumar2018non}). Further, $l = \sqrt{g(g - 2\gamma^{AD})}$, where the coupling strength $\gamma^{AD}$ of AD noise is related to the qubit relaxation time ($\tau_s = \frac{1}{\gamma^{AD}}$), and $g$ is the line width that depends on the reservoir correlation time ($\tau_r = \frac{1}{g}$). The system exhibits Markovian and non-Markovian evolution of a state if $2\gamma^{AD} \ll g$ and $2\gamma^{AD} \gg g $, respectively, for the AD noise~\cite{Bellomo2007NMAD}.

% \subsection{Fidelity of noisy negative quantum states}
\begin{figure}[H]
    \centering
    \includegraphics[width=1\columnwidth]{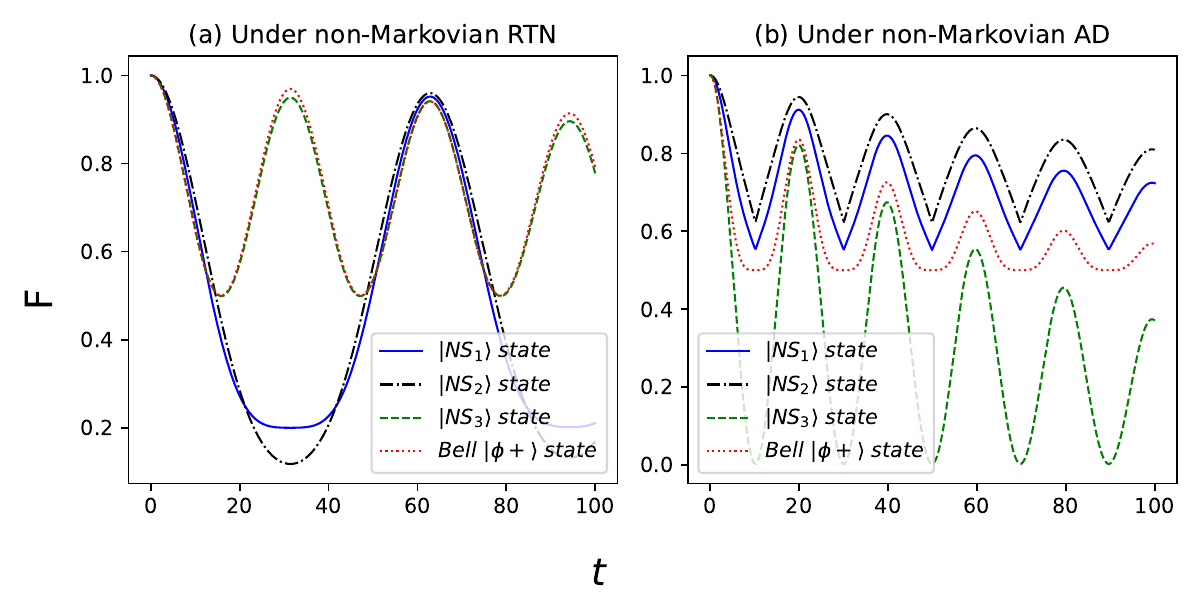}
    \caption{Variation of fidelity of $\ket{NS_1}$, $\ket{NS_2}$, $\ket{NS_3}$ and $\ket{\phi^{+}}$ Bell state with time (a) under non-Markovian RTN for $b = 0.05$ and $\gamma^{RTN} = 0.001$, (b) under non-Markovian AD for $g = 0.01$ and $\gamma^{AD} = 5$.}
    \label{Fidelity_NMRTN_NMAD}
\end{figure}
The variation of fidelity $F$, between the original state and the state after applying the non-Markovian noise, of two-qubit negative quantum states and the Bell state under non-Markovian RTN and AD noise is illustrated in Fig.~\ref{Fidelity_NMRTN_NMAD}. We can observe that, in the presence of non-Markovian RTN noise, the $\ket{NS_3}$ state and the Bell state's fidelity behave equivalently, while under non-Markovian AD noise, the $\ket{NS_2}$ state is the most resilient among all the other considered states.

As part of exploring the applications of two-qubit negative quantum states, we study their role in enhancing parameter estimation sensitivity through quantum Fisher information. Also, we investigate the optimal CHSH inequality violation and present a teleportation circuit based on $\ket{NS_3^{\prime\prime}}$ state, further demonstrating their potential in quantum information processing and communication.

\subsection{Quantum Fisher information}
Quantum Fisher information (QFI) is the central quantity in quantum metrology~\cite{Bollinger1996Optimalfrequency,peters1999measurement}. It quantifies the ultimate precision limit with which a parameter can be estimated and connected to the lower bound on the variance of an unbiased parameter via the quantum Cram\'er-Rao bound~\cite{helstrom1969quantum}.

For a density matrix $\rho(\phi)$, with a parameter $\phi$ acquired by an $SU(2)$ rotation, i.e., $\rho_{\phi} = U_{\phi} \rho U_{\phi}^{\dag}$, where $U_{\phi} = e^{i \phi J_{\vec{n}}}$ with 
\begin{equation}
    J_{\vec{n}} = \sum_{\alpha = x,y,z}\frac{1}{2}n_{\alpha}\sigma_{\alpha},
\end{equation}
being the angular momentum operator in the $\vec{n}$ direction, and $\sigma_{\alpha}$ are the Pauli matrices. The QFI, for an unbiased estimator $\langle \hat{\phi} \rangle = \phi$, is defined as~\cite{ma2011quantum}
\begin{equation}
    F = Tr[\rho(\phi) L_{\phi}^2],
    \label{QFI}
\end{equation}
where $L_{\phi}$ is the symmetric logarithmic derivative determined by the following equation,
\begin{equation}
    \frac{\partial}{\partial \phi}\rho(\phi) = \frac{1}{2}[\rho(\phi)L_{\phi} + L_{\phi}\rho(\phi)].
    \label{derivative_L_phi}
\end{equation}
Now, using Eq. (\ref{derivative_L_phi}), explicit form of $L_{\phi}$ can be derived and the QFI for a density matrix $\rho$ is thus given as~\cite{ma2011quantum},

\begin{equation}
    F(\rho, J_{\vec{n}}) = \sum_{i \neq j} \frac{2(p_i - p_j)^2}{p_i + p_j}|\langle i|J_{\vec{n}}|j\rangle|^2 = \vec{n}~\mathcal{C}~\vec{n}^T,
\end{equation}
where $\vec{n}$ is a normalized three-dimensional vector. Further, \{$p_i$, $p_j$\} and \{$\ket{i}$, $\ket{j}$\} are the eigenvalues and eigenvectors of $\rho$, respectively. Further, the matrix elements of the symmetric matrix $\mathcal{C}$ can be obtained as 
\begin{equation}
    \mathcal{C}_{kl} = \sum_{i \neq j}\frac{(p_i - p_j)^2}{p_i + p_j}[\langle i|J_k|j\rangle \langle j|J_l|i\rangle + \langle i|J_l|j\rangle \langle j|J_k|i\rangle].
\end{equation}

The QFI, $F(\rho, J_{\vec{n}})$, measures the sensitivity of the state concerning the rotations along the $\vec{n}$ direction. Additionally, the maximal mean QFI is defined as follows~\cite{hyllus2012fisher},
\begin{equation}
    {\bar{F}}_{max} = \frac{1}{N} \underset{\vec{n}}{\max} F(\rho, J_{\vec{n}}) = \frac{\lambda_{max}}{N},
\end{equation}
where $\lambda_{max}$ is the largest eigenvalue of the symmetric matrix $\mathcal{C}$ and $N$ is the number of two-level particles. The ${\bar{F}}_{max}$ characterizes the phase sensitivity, i.e., the sensitivity of a state with respect to $SU(2)$ rotations, and is independent of the $\vec{n}$. Now, we calculate the ${\bar{F}}_{max}$ of two-qubit negative quantum states and the Bell state in the presence of the non-Markovian AD noise to study the effect of decoherence on their phase sensitivity. Figure \ref{NS1_NS2_fisher_info_NMAD} shows the variation of the ratio of maximal mean QFI for $\ket{NS_i}$ state to that of $\ket{\phi^{+}}$ state, denoted by $\zeta_{\ket{NS_i}}$, under non-Markovian AD noise. In Fig. \ref{NS1_NS2_fisher_info_NMAD}, wherever $\zeta_{\ket{NS_i}} > 1$, we can observe that ${\bar{F}}_{max}$ for $\ket{NS_i}$ state is greater than ${\bar{F}}_{max}$ for $\ket{\phi^{+}}$ state. Figure \ref{NS1_NS2_fisher_info_NMAD} shows that  $\ket{NS_1}$ and $\ket{NS_2}$ states maintain a higher ${\bar{F}}_{max}$ for longer duration in comparison to the Bell $\ket{\phi^{+}}$ state making them better suited for realistic quantum metrology applications under noise. Because the higher maximal mean QFI leads to better phase estimation via the Cram\'er-Rao bound. 
\begin{figure}[h]
    \centering
    \includegraphics[width=1\columnwidth]{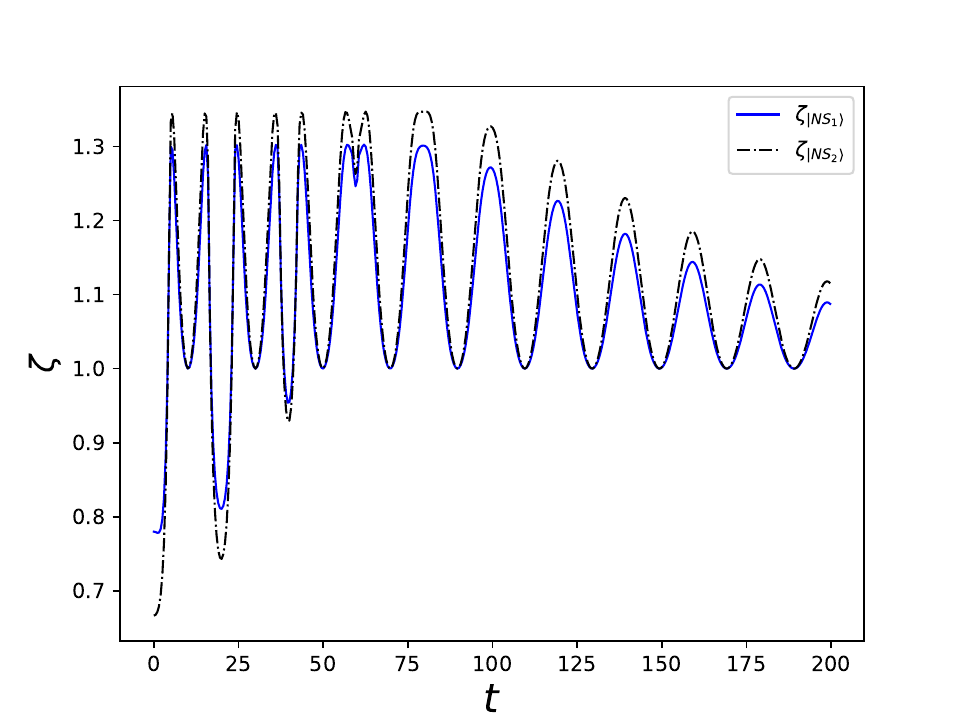}
    \caption{Variation of $\zeta_{\ket{NS_1}}$ and $\zeta_{\ket{NS_2}}$ with respect to time under non-Markovian AD for $g = 0.01$ and $\gamma^{AD} = 5$.}
    \label{NS1_NS2_fisher_info_NMAD}
\end{figure}
\subsection{Violation of the CHSH inequality} \label{sec:Optimal_CHSH_inequality_violation}
The optimal violation of the Clauser-Horne-Shimony-Holt (CHSH) inequality constitutes a fundamental benchmark for detecting the quantum nonlocality of a general two-qubit state. Given a two-qubit density matrix $\rho$, the maximal CHSH violation is determined through the spectral properties of matrix $T_{\rho}^{T}T_{\rho}$ as follows~\cite{HORODECKI1995340}, 
\begin{equation}
    S_{max} = 2\sqrt{\lambda_1 + \lambda_2},
    \label{S_max_eq.}
\end{equation}
where $\lambda_1$ and $\lambda_2$ are the largest two eigenvalues of the $T_{\rho}^{T}T_{\rho}$ matrix. The elements of $3 \times 3$ correlation matrix $T_{\rho}$ are defined as $t_{ij} = {\rm Tr}[\rho. (\sigma_i \otimes \sigma_j)]$, where $\sigma_1$, $\sigma_2$, and $\sigma_3$ are the Pauli matrices. We study the optimal CHSH inequality violation of the two-qubit negative quantum states and all the Bell states with weak measurement (WM) and quantum measurement reversal (QMR) detailed in Appendix~\ref{WM_QMR} in the presence of non-Markovian AD noise. Figure~\ref{NS2_CHSH_maximal_violation_NMAD_with_WM} depicts the variation of the ratio of $S_{max}$ for $\ket{NS_i}$ state to the $S_{max}$ for $\ket{\phi^{+}}$ state with WM and QMR, denoted by $\eta_{\ket{NS_i}}$, particularly for the state which perform best among all, i.e., $\ket{NS_2}$ state. We observe that the $\ket{NS_2}$ state significantly outperforms the Bell $\ket{\phi^{+}}$ state in preserving CHSH violation over time. The $\ket{NS_2}$ state shows more frequent revivals above the classical limit $2$, which means it retains non-locality better, while the Bell $\ket{\phi^{+}}$ state often stays below $2$ under non-Markovian AD noise.
\begin{figure}[h]
    \centering
    \includegraphics[width=1\columnwidth]{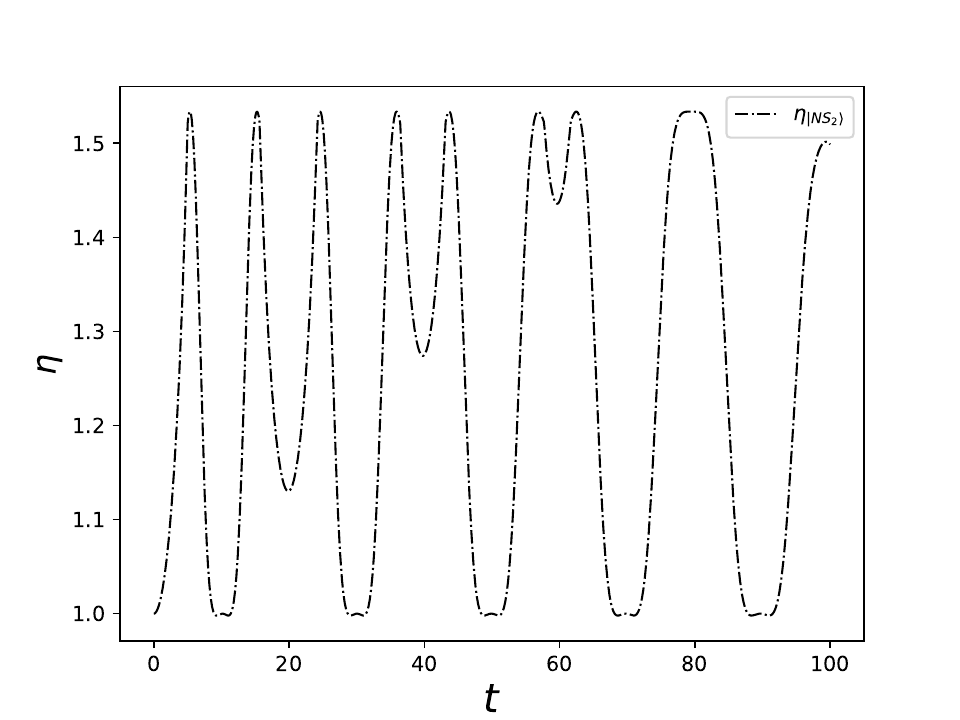}
    \caption{The temporal variation of $\eta_{\ket{NS_2}}$, specifically the ratio $S_{max \ket{NS_2}}/S_{max \ket{\phi^{+}}}$ under non-Markovian AD channel with WM and QMR for $\ket{NS_2}$ ($p = 0.05$, $q = 0.74$), and for Bell $\ket{\phi^{+}}$ state ($p = 0.05$, $q = 0.05$). The non-Markovian AD channel parameters are $g = 0.01$ and $\gamma^{AD} = 5$.}
    \label{NS2_CHSH_maximal_violation_NMAD_with_WM}
\end{figure}
Further, a detailed methodology and comprehensive analysis for identifying optimal states, among the Bell states and two-qubit negative quantum states, that facilitate universal quantum teleportation is provided in Appendix~\ref{WM_QMR}.
\subsection{Quantum teleportation using $\ket{NS_3^{\prime\prime}}$} \label{sec:T_circuit}
\begin{figure}[h]
    \centering
    \includegraphics[height=25mm,width=0.5\textwidth]{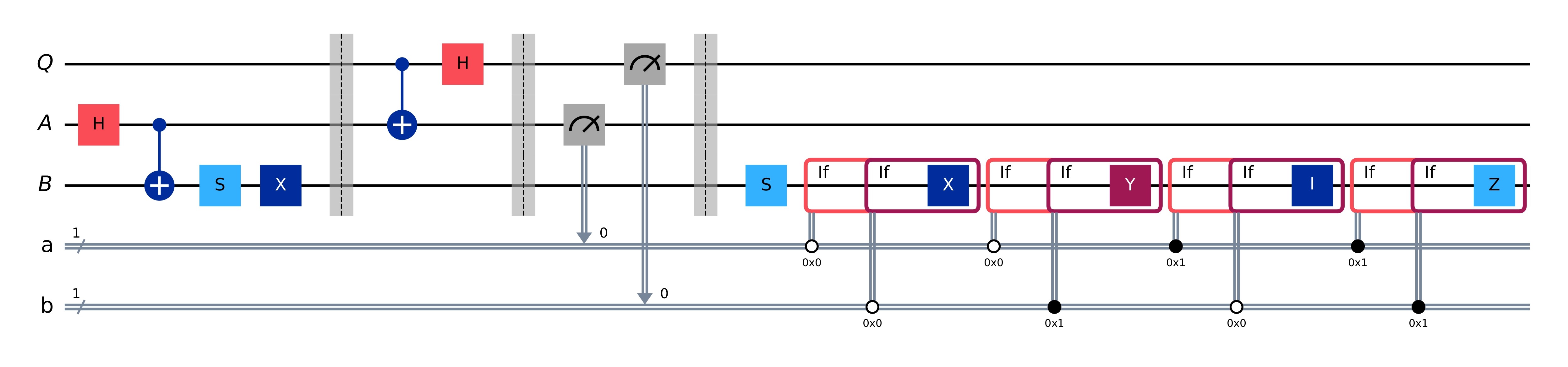}
    \caption{Circuit for implementing quantum teleportation scheme using $\ket{NS_3^{\prime\prime}}$ as an entangled resource. The quantum gates in the blocks annotated with ``if" are applied conditioned on the values of the classical bits corresponding to the measurement outcomes.}
    \label{Teleportation_NS3_double_prime}
\end{figure} 
\begin{figure}
    \includegraphics[width=1\columnwidth]{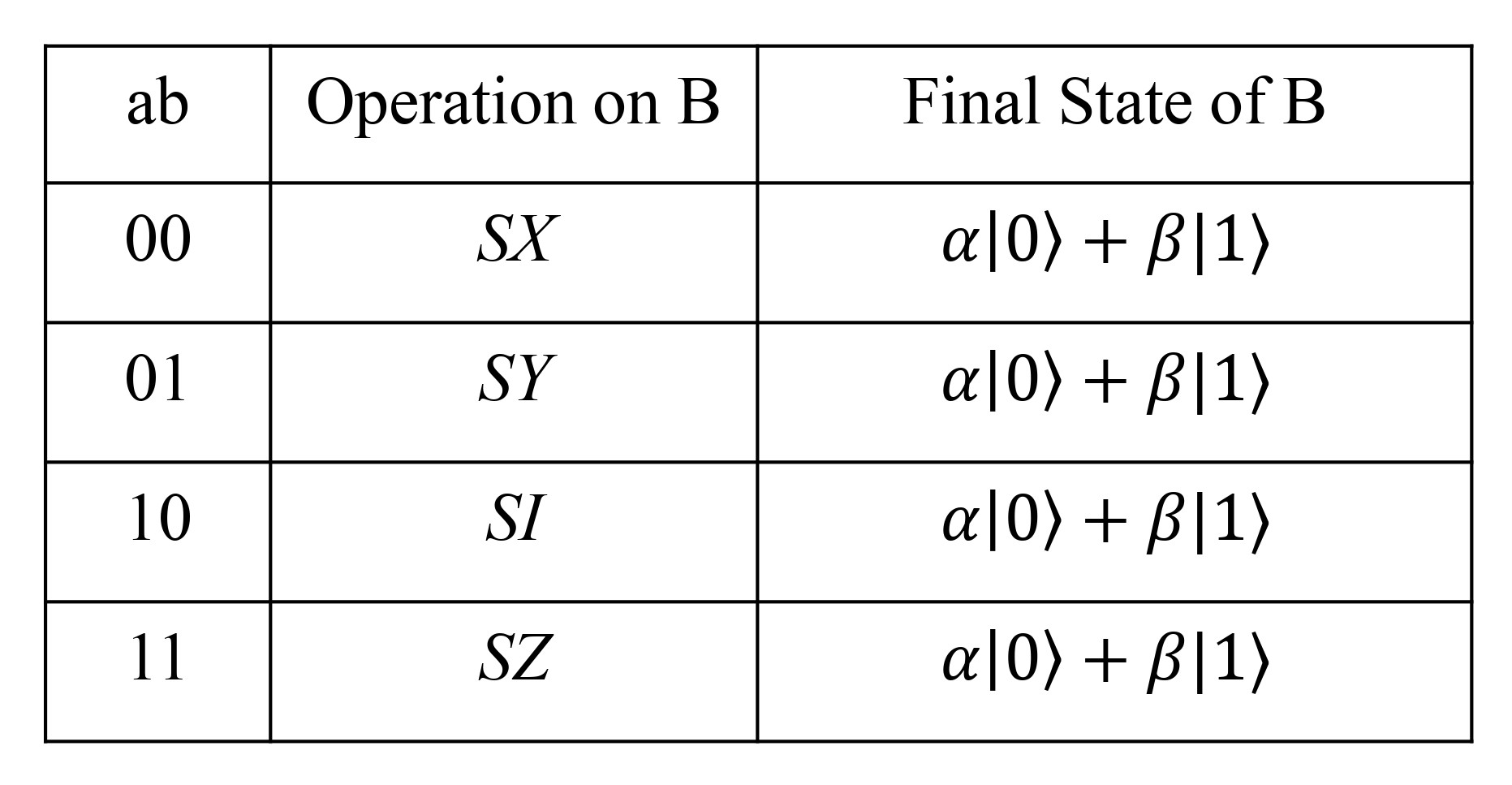}
    \captionsetup{labelformat=empty}
    \caption{TABLE II: Alice transmits the classical information (ab) to Bob through a classical communication channel. Based on the received classical bits, Bob performs specific unitary operations on his qubit (denoted as operation on $B$). Following the application of these operations, the resulting final state of Bob's qubit (Final state of $B$) corresponds precisely to the unknown quantum information originally intended for teleportation.}
    \label{Teleportation_table}
 \end{figure}
Quantum teleportation is a method used to transfer quantum information from one end to another, even when no direct quantum communications channel connects them~\cite{Teleporting1993}. To accomplish this, the sender, referred to as ``Alice," and the receiver, referred to as ``Bob," must make prior arrangements for a shared pair of Einstein-Podolsky-Rosen (EPR) correlated particles. Alice performs the Bell measurement on her EPR particle and the unknown quantum state. She then transmits the classical outcome of this measurement to Bob. With this knowledge, Bob can transform the state of his EPR particle into an identical copy of the unknown quantum state~\cite{nielsen2010quantum}. The teleportation protocol using the Bell state as an EPR pair has been extensively studied. Since the two-qubit $\ket{NS_3^{\prime\prime}}$ state also has the maximum concurrence, i.e., 1, we now introduce the teleportation protocol using this state as a shared EPR pair. The corresponding quantum teleportation circuit is shown in Fig~\ref{Teleportation_NS3_double_prime}. ``A" (Alice) denotes the sender and ``B" (Bob) the receiver, who share the entangled $\ket{NS3^{\prime\prime}}$ state in Fig.~\ref{Teleportation_NS3_double_prime}. ``Q" represents the unknown qubit to be teleported from Alice to Bob. Following a Bell-state measurement, Alice transmits two classical bits, ``a" and ``b," to Bob. The corrective operations that Bob must apply, determined by the values of ``a" and ``b," are specified by the conditional statements and summarized in Table~II. After applying the appropriate operations, Bob's resulting state for an initial unknown state $\alpha\ket{0} + \beta\ket{1}$ is also listed in Table~II. The teleportation circuit illustrated in Fig.~\ref{Teleportation_NS3_double_prime} is verified on the Qiskit~\cite{qiskit2024} by successfully teleporting random quantum information from Alice to Bob. Moreover, the teleportation protocol can be executed utilizing an alternative $\ket{NS_3}$ and $\ket{NS_3^{\prime}}$ states, as they possess entanglement at the maximal likelihood of the Bell state. Nonetheless, in this instance, identifying the unitary transformations that Bob must implement is challenging. 
\section{Conclusions}\label{sec:conclusion}
In this work, we proposed methods for preparing stable entangled states, specifically focusing on two-qubit negative quantum states. We presented explicit quantum circuits to facilitate their implementation. Employing ideal and error-mitigated quantum state tomography on IBM's real quantum hardware and simulator, we validated the preparation of these states. We achieved fidelities equivalent to those of an ideal Bell state. The resilience of these states against (non)-Markovian noise is demonstrated through detailed analyses of fidelity estimation, maximal mean quantum Fisher information, optimal CHSH inequality violation, and performance in universal quantum teleportation, both with and without the application of weak measurement (WM) and quantum measurement reversal (QMR). These findings emphasize the practical relevance of negative quantum states for quantum sensing and metrology, where maintaining precision over extended timescales is critical. Furthermore, the application of WM and QMR techniques effectively extends the coherence time of these states, enhancing their viability in quantum computing, quantum key distribution, quantum teleportation, and superdense coding—areas where the preservation of quantum correlations is critical. Future directions include the development of fault-tolerant circuits for the realization of the two-qubit negative states and the generalization of these states to multiqubit architectures to design resilient quantum memories.

\section{Acknowledgements}
J. L. acknowledges Prof. Anirban Pathak, Devvrat Tiwari, and Vivek B. Sabale for their useful discussions and valuable insights.
\bibliography{BibTexfile}
\bibliographystyle{apsrev4-2}
\clearpage
\onecolumngrid
\appendix
\section{\label{Unitary transformations} Unitary transformations of negative quantum states and their corresponding quantum circuits}
\begin{figure}[H]
    \centering
    \includegraphics[height=85mm,width=0.85\textwidth]{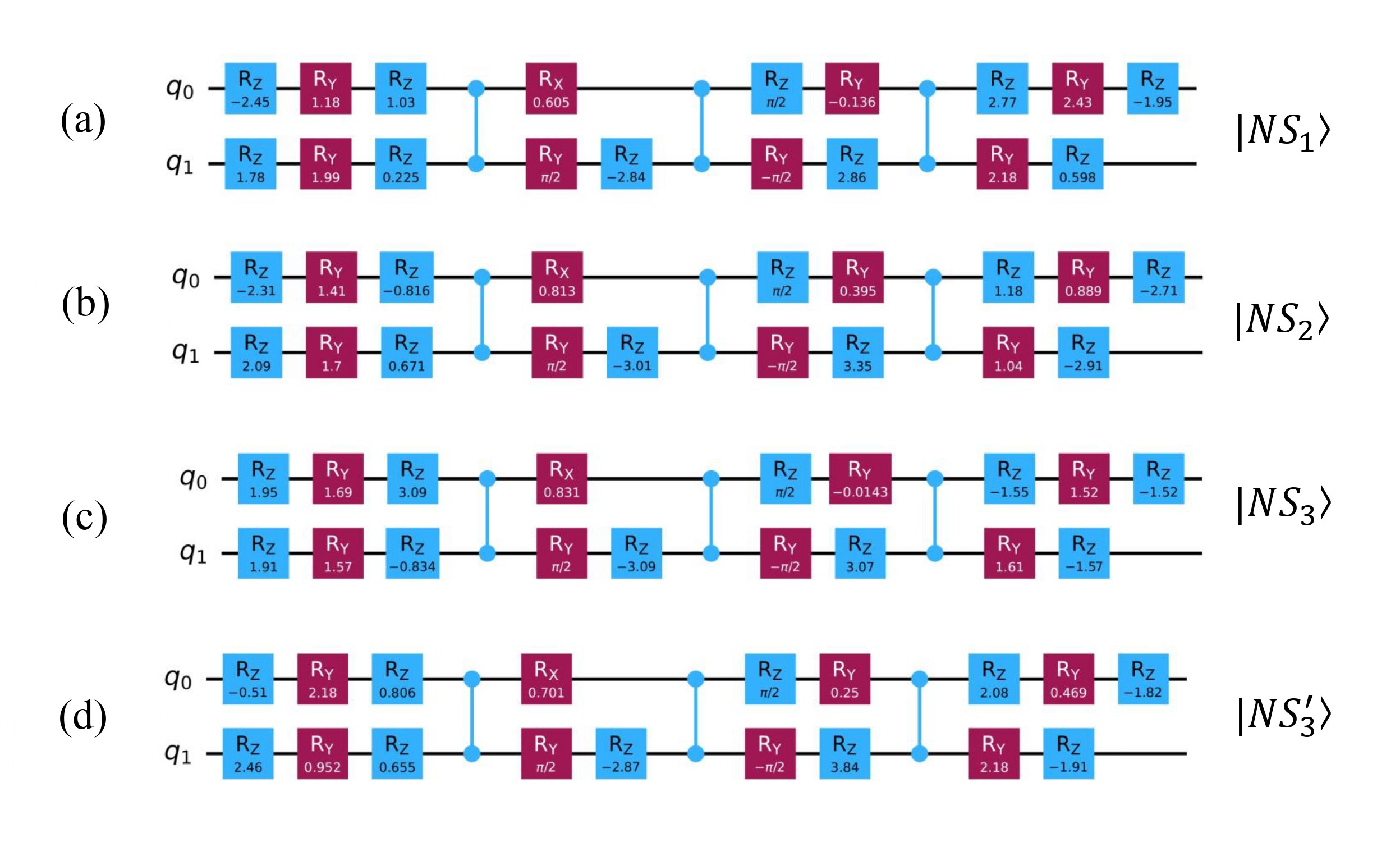}
    \caption{Quantum circuits to generate the two-qubit $\ket{NS_1}$, $\ket{NS_2}$, $\ket{NS_3}$, and $\ket{NS_3^{\prime}}$ states from the $\ket{00}$ state using $H$, $R_x$, $R_z$, and $CZ$ gates are shown in subfigures (a), (b), (c), and (d) respectively. Here, $q_0$ and $q_1$ represent the qubits in the $\ket{00}$ state.}
    \label{NS1_NS2_NS3__NS3_prime_circuit}
\end{figure}
The unitary transformation matrices obtained using the Gram-Schmidt procedure mentioned in the main text for realizing $\ket{NS_1}$, $\ket{NS_2}$, $\ket{NS_3}$, $\ket{NS_3^{\prime}}$, and $\ket{NS_3^{\prime\prime}}$ states from the $\ket{00}$ state are given as follows:
    \begin{equation}
    U_{NS_1} =\left(
    \begin{array}{cccc}
     -0.742977+0. i & 0.669317\, +0. i & 0.\, +0. i & 0.\, +0. i \\
     -0.357599+0.357599 i & -0.396953+0.396953 i & 0.655059\, +0. i & 0.\, +0. i \\
     0.101586\, +0.101586 i & 0.112766\, +0.112766 i & 0.\, +0.247581 i & 0.944792\, +0. i \\
     -0.414237+0. i & -0.459824+0. i & -0.504778-0.504778 i & 0.231698\, -0.231698 i \\
    \end{array}
    \right)
    \end{equation}
    
    \begin{equation}
        U_{NS_2} = \left(
    \begin{array}{cccc}
     0.788675\, +0. i & 0.61481\, +0. i & 0.\, +0. i & 0.\, +0. i \\
     -0.288675+0.288675 i & 0.370311\, -0.370311 i & 0.747712\, +0. i & 0.\, +0. i \\
     -0.288675-0.288675 i & 0.370311\, +0.370311 i & 0.\, -0.589702 i & 0.459701\, +0. i \\
     -0.211325+0. i & 0.271086\, +0. i & -0.215846-0.215846 i & -0.627963+0.627963 i \\
    \end{array}
    \right)
    \end{equation}
    
    \begin{equation}
        U_{NS_3} = \left(
    \begin{array}{cccc}
     -0.0508479+0. i & 0.998706\, +0. i & 0.\, +0. i & 0.\, +0. i \\
     0.631483\, -0.228733 i & 0.0321511\, -0.0116456 i & 0.740096\, +0. i & 0.\, +0. i \\
     -0.27958-0.68233 i & -0.0142345-0.03474 i & 0.0277425\, +0.670334 i & 0.0687934\, +0. i \\
     0.0508479\, +0. i & 0.00258886\, +0. i & -0.0434981-0.0157557 i & 0.378252\, -0.923143 i \\
    \end{array}
    \right)
    \end{equation}

    \begin{equation}
        U_{NS_3^{\prime}} = \left(
        \begin{array}{cccc}
         -0.575107+0. i & 0.818078\, +0. i & 0.\, +0. i & 0.\, +0. i \\
         -0.345634+0.310025 i & -0.242979+0.217946 i & 0.823336\, +0. i & 0.\, +0. i \\
         -0.265082-0.229473 i & -0.186352-0.161319 i & -0.0371656-0.293085 i & 0.85384\, +0. i \\
         0.575107\, +0. i & 0.404298\, +0. i & 0.360743\, +0.323577 i & 0.393558\, -0.34069 i \\
        \end{array}
        \right)
    \end{equation}

    \begin{equation}
        U_{NS_3^{\prime\prime}} = \left(
    \begin{array}{cccc}
     0.\, +0. i & 0.\, +0. i & 0.\, +0.707106781i & 0.\, +0.707106781i \\
     0.\, +0.707106781i & 0.\, -0.707106781i & 0.\, +0. i & 0.\, +0. i \\
     +0.707106781+0. i & +0.707106781+0. i & 0.\, +0. i & 0.\, +0. i \\
     0.\, +0. i & 0.\, +0. i & +0.707106781+0. i & -0.707106781+0. i \\
    \end{array}
    \right)
    \end{equation}
The quantum circuits generated by Qiskit~\cite{qiskit2024}, corresponding to the unitary matrices above, are illustrated in Fig. \ref{NS1_NS2_NS3__NS3_prime_circuit}.

\section{\label{WM_QMR} Optimal CHSH inequality violation, Concurrence, and Teleportation fidelity under non-Markovian AD noise with(without) weak measurement}
\begin{figure}[h]
    \centering
    \includegraphics[width=1\columnwidth]{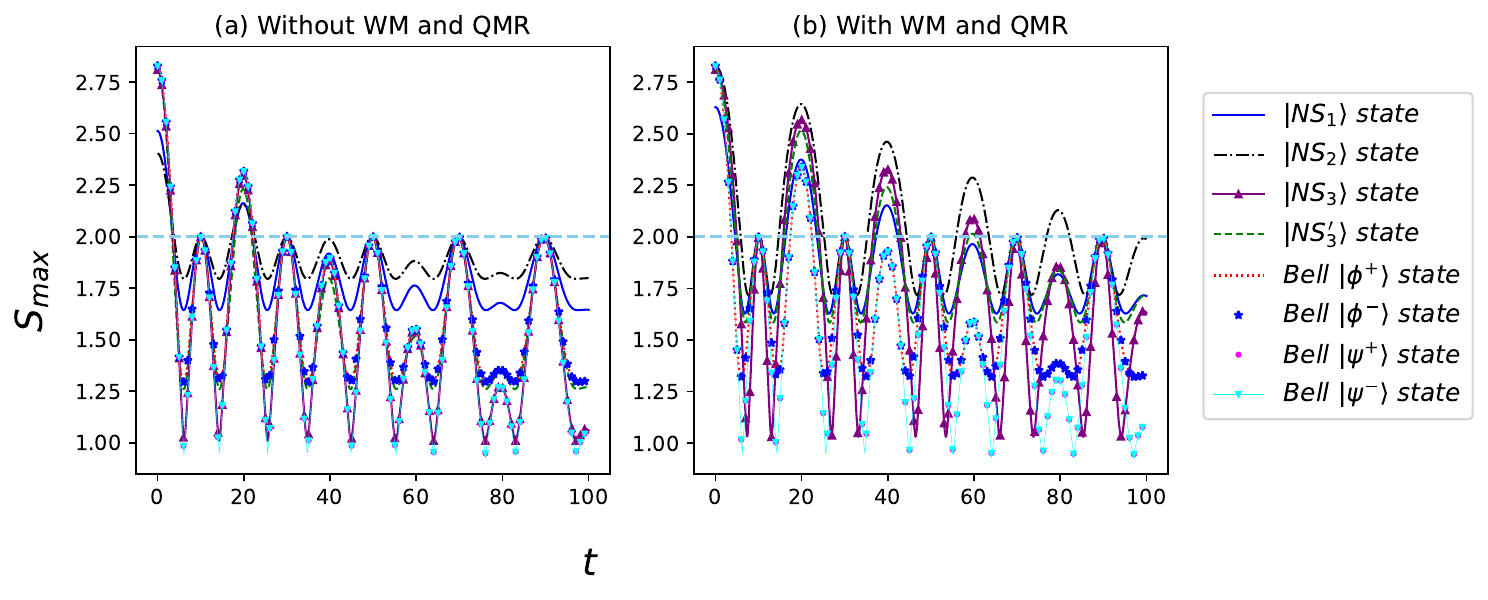}
    \caption{Variation of optimal CHSH inequality violation of $\ket{NS_1}$, $\ket{NS_2}$, $\ket{NS_3}$, $\ket{NS_3^{\prime}}$ and all the Bell states under non-Markovian AD channel without WM and QMR in subplot (a) and with WM and QMR in subplot (b) with time. Here, for $\ket{NS_1}$ ($p = 0.17$, $q = 0.54$), for $\ket{NS_2}$ ($p = 0.05$, $q = 0.74$), for $\ket{NS_3}$ ($p = 0.54$, $q = 0.54$), for $\ket{NS_3^{\prime}}$ ($p = 0.58$, $q = 0.58$), for Bell $\ket{\phi^{+}}$, $\ket{\phi^{-}}$ states ($p = 0.05$, $q = 0.05$) and for Bell $\ket{\psi^{+}}$, $\ket{\psi^{-}}$ states ($p = 0.01$, $q = 0.05$). The non-Markovian AD channel parameters are $g = 0.01$ and $\gamma^{AD} = 5$.}
    \label{smax_NMAD}
\end{figure}
Unlike traditional von Neumann measurement, the weak measurement (WM) retrieves information from the system without inducing its collapse into an eigenstate. An appropriate quantum measurement reversal (QMR) can reconstruct the state with a specific success probability. Consequently, it has been demonstrated that WM and QMR can augment and safeguard quantum correlations in qubit and qutrit systems against non-Markovian noise~\cite{he2020enhancing, Lalita_2024ProtectingQC}. Additionally, photonic and superconducting quantum systems can be used for the experimental implementation of WM and QMR~\cite{kim2009reversing}. The physical framework for preserving quantum correlations of two-qubit entangled states using WM and QMR is as follows. Initially, Charlie applies a weak measurement ${M}_{\textit{WM}}(w_1, w_2)$, as defined in Eq. (\ref{wm}), to the entangled state before its transmission through non-Markovian noisy quantum channels to Alice and Bob. Upon receiving their respective qubits, Alice and Bob perform quantum measurement reversal ${M}_{\textit{QMR}}(wr_1, wr_2)$, as specified in Eq. (\ref{qmr}). By appropriately tuning the parameters $(w_1, w_2)$ and $(wr_1, wr_2)$, the final state ${\rho}_f(t)$, given in Eq. (\ref{eq:WM_QMR}), can be restored to a maximally entangled form. The non-unitary weak WM and QMR operations, characterized by WM and QMR strength parameters $(w_1, w_2)$ and $(wr_1, wr_2)$, respectively, are defined as follows:
\begin{equation}
   \begin{aligned}
     {M}_{\textit{WM}}(w_1, w_2) = \begin{pmatrix}
                                1 & 0\\
                                0 & \sqrt{1-w_1}
                               \end{pmatrix} \otimes \begin{pmatrix}
                                1 & 0\\
                                0 & \sqrt{1-w_2}
                               \end{pmatrix},
      \end{aligned}
      \label{wm}
\end{equation}
\begin{equation}
   \begin{aligned}
     {M}_{\textit{QMR}}(wr_1, wr_2) = \begin{pmatrix}
                                \sqrt{1-wr_1} & 0\\
                                0 & 1
                               \end{pmatrix} \otimes \begin{pmatrix}
                                \sqrt{1-wr_2} & 0\\
                                0 & 1
                               \end{pmatrix}.
      \end{aligned}
      \label{qmr}
\end{equation}
The resulting final state after the sequential application of WM, evolution through the non-Markovian channel Kraus operators (${K}_{ij}$), and QMR is
\begin{equation}
 {\rho}_f(t) = \frac{{M}_{\textit{QMR}}\left( \sum_{i = 0}^{1}\sum_{j = 0}^{1} {K}_{ij}[{M}_{\textit{WM}} \rho_0 {M}_{\textit{WM}}^{\dag}] {K}_{ij}^{\dag}\right) {M}_{\textit{QMR}}^{\dag}}{P^{succ}},
 \label{eq:WM_QMR}
\end{equation}
here $\rho_0$ is the two-qubit entangled state at $t = 0$ and as the WM and QMR are probabilistic in nature $P^{succ} = Tr[{M}_{\textit{QMR}}\left( \sum_{i = 0}^{1}\sum_{j = 0}^{1} {K}_{ij}[{M}_{\textit{WM}}{\rho}(0){M}_{\textit{WM}}^{\dag}] {K}_{ij}^{\dag}\right) {M}_{\textit{QMR}}^{\dag}]$ is its success probability~\cite{he2020enhancing, Lalita_2024ProtectingQC}. In~\cite{Lalita_2024ProtectingQC}, $\ket{\phi^{+}}$ Bell state and the two-qubit negative quantum states are taken into account to find the optimal state for universal quantum teleportation in the presence of non-Markovian RTN and AD noise with(without) weak measurement (WM)~\cite{oreshkov2005weak, katz2008reversal, breuer2002theory}. Here, we extend this study to find the most suitable candidate for optimal CHSH inequality violation and universal quantum teleportation among all the four Bell states (given below) and two-qubit negative quantum states in the presence of non-Markovian AD noise.
\begin{eqnarray}
    \begin{aligned}
        \ket{\phi^{+}} &= {1/\sqrt{2}}\left(~1~~0~~0~~1~\right)^T;
        \ket{\phi^{-}} = {1/\sqrt{2}}\left(~1~~0~~0~~{-1}~\right)^T; \nonumber \\
        \ket{\psi^{+}} &= {1/\sqrt{2}}\left(~0~~1~~1~~0~\right)^T; \nonumber
        \ket{\psi^{-}} = {1/\sqrt{2}}\left(~0~~1~~{-1}~~0~\right)^T. 
    \end{aligned}
    \label{Bell_states}
\end{eqnarray}
\begin{figure}
    \centering
    \includegraphics[width=1\columnwidth]{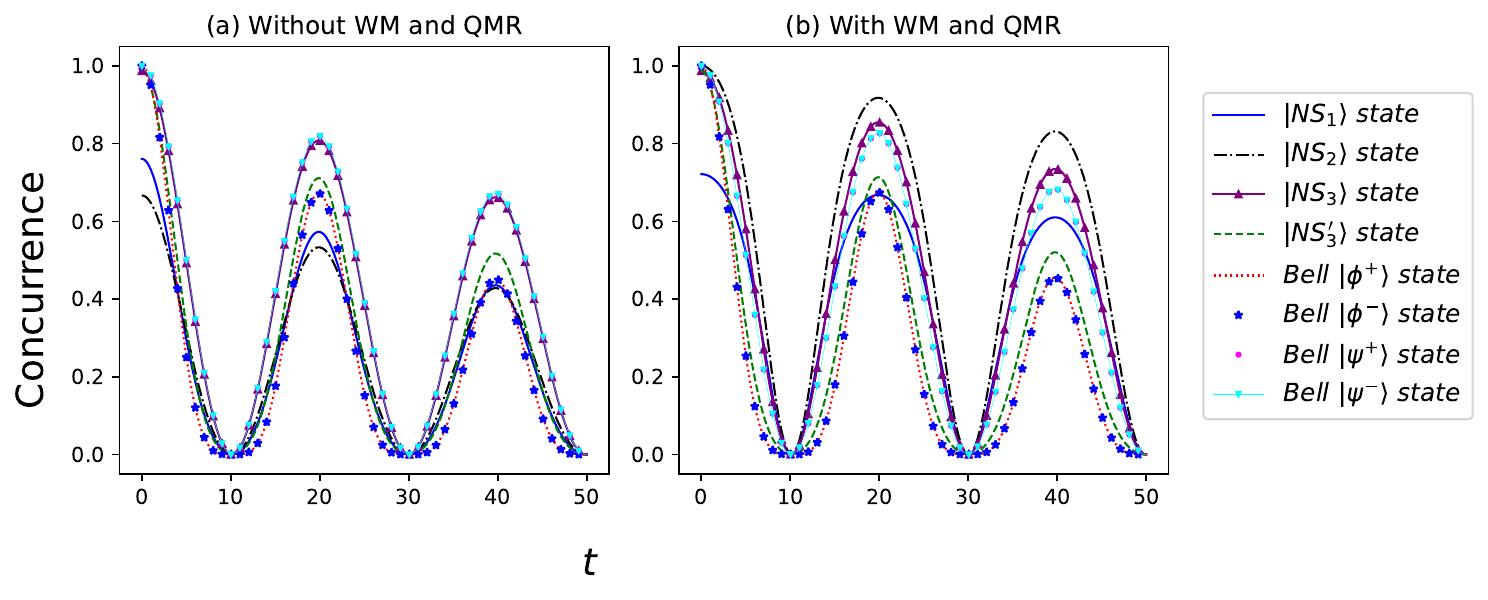}
    \caption{Variation of concurrence of $\ket{NS_1}$, $\ket{NS_2}$, $\ket{NS_3}$, $\ket{NS_3^{\prime}}$ and all the Bell states under non-Markovian AD channel without WM and QMR in subplot (a) and with WM and QMR in subplot (b) with time. Here, for $\ket{NS_1}$ ($p = 0.17$, $q = 0.54$), for $\ket{NS_2}$ ($p = 0.05$, $q = 0.74$), for $\ket{NS_3}$ ($p = 0.05$, $q = 0.05$), for $\ket{NS_3^{\prime}}$ ($p = 0.3$, $q = 0.3$), for Bell $\ket{\phi^{+}}$, $\ket{\phi^{-}}$ states ($p = 0.01$, $q = 0.01$) and for Bell $\ket{\psi^{+}}$, $\ket{\psi^{-}}$ states ($p = 0.01$, $q = 0.05$). The non-Markovian AD channel parameters are $g = 0.01$ and $\gamma^{AD} = 5$.}
    \label{concur_NMAD}
\end{figure}%
\begin{figure}
    \centering
    \includegraphics[width=1\columnwidth]{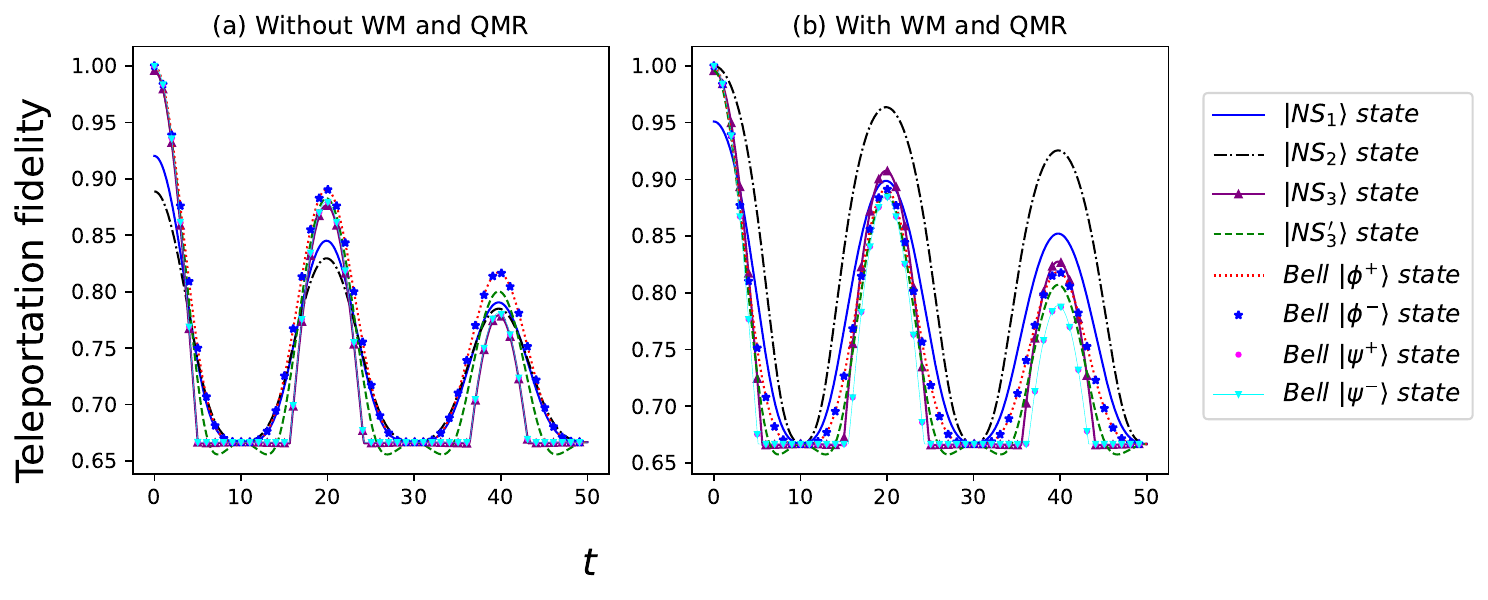}
    \caption{Variation of teleportation fidelity of $\ket{NS_1}$, $\ket{NS_2}$, $\ket{NS_3}$, and Bell state under non-Markovian AD channel without WM and QMR in subplot (a), and with WM and QMR in subplot (b) with time. Here, for $\ket{NS_1}$ ($p = 0.17$, $q = 0.54$), for $\ket{NS_2}$ ($p = 0.05$, $q = 0.74$), for $\ket{NS_3}$ ($p = 0.05$, $q = 0.05$), for $\ket{NS_3^{\prime}}$ ($p = 0.3$, $q = 0.3$), for Bell $\ket{\phi^{+}}$, $\ket{\phi^{-}}$ states ($p = 0.01$, $q = 0.01$) and for Bell $\ket{\psi^{+}}$, $\ket{\psi^{-}}$ states ($p = 0.01$, $q = 0.05$). The non-Markovian AD channel parameters are $g = 0.01$ and $\gamma^{AD} = 5$.}
    \label{TFid_NMAD}
\end{figure}
\begin{figure}
    \centering
    \includegraphics[width=1\columnwidth]{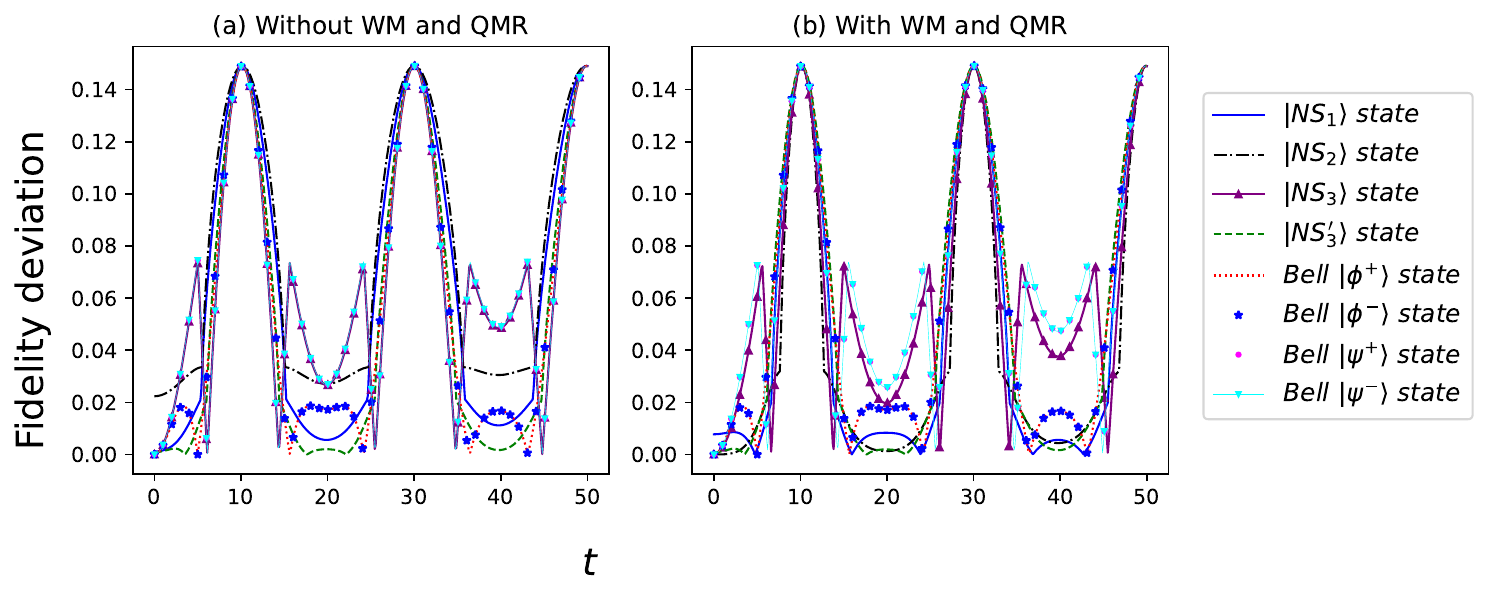}
    \caption{Variation of fidelity deviation of $\ket{NS_1}$, $\ket{NS_2}$, $\ket{NS_3}$, and Bell state under non-Markovian AD channel without WM and QMR in subplot (a), and with WM and QMR in subplot (b) with time. Here, for $\ket{NS_1}$ ($p = 0.17$, $q = 0.54$), for $\ket{NS_2}$ ($p = 0.05$, $q = 0.74$), for $\ket{NS_3}$ ($p = 0.05$, $q = 0.05$), for optimal $NS3$ ($p = 0.3$, $q = 0.3$), for Bell $\ket{\phi^{+}}$, $\ket{\phi^{-}}$ states ($p = 0.01$, $q = 0.01$) and for Bell $\ket{\psi^{+}}$, $\ket{\psi^{-}}$ states ($p = 0.01$, $q = 0.05$). The non-Markovian AD channel parameters are $g = 0.01$ and $\gamma^{AD} = 5$.}
    \label{FD_NMAD}
\end{figure}

The optimal CHSH inequality violation, denoted by $S_{max}$, Eq.~(\ref{S_max_eq.}) in the main text, is computed using the final state ${\rho}_f(t)$ obtained after sequential application of WM, non-Markovian amplitude damping (AD) noise, and QMR for all Bell states and selected two-qubit negative quantum states. The temporal evolution of $S_{max}$ under the influence of WM and QMR in the presence of non-Markovian AD noise is depicted in Fig.~\ref{smax_NMAD}(b). In contrast, Fig.~\ref{smax_NMAD}(a) depicts the temporal behavior of $S_{max}$ for the same set of states when only non-Markovian AD noise is present, without WM and QMR interventions.
From Fig.~\ref{smax_NMAD}, it is evident that in the absence of WM and QMR, the states $|NS_1\rangle$ and $|NS_2\rangle$ do not achieve maximal CHSH inequality violation (i.e., $2\sqrt{2}$), yet they exhibit slower decoherence compared to the states that initially achieve maximal violation. However, when WM and QMR are applied, the $|NS_2\rangle$ state attains maximal CHSH inequality violation and displays more frequent and pronounced random fluctuations above the classical bound of $2$ than any other considered state. Furthermore, the $|NS_3\rangle$ and $|NS_3'\rangle$ states also consistently exhibit stronger nonlocal correlations over extended durations relative to the Bell states.

Additionally, building upon our earlier work~\cite{Lalita_2024ProtectingQC}, we have expanded our analysis to identify the optimal state for achieving universal quantum teleportation by considering all Bell states and the aforementioned two-qubit entangled states. Quantifying the entanglement content is critical for universal quantum teleportation. Thus, we calculate concurrence~\cite{wootters1998entanglement}, a prominent measure of entanglement, for all the states considered. The variation of concurrence for the above Bell states and the negative quantum states under non-Markovian AD noise with(without) weak measurement is shown in Fig.~\ref{concur_NMAD}. We observe that the $\ket{NS_3}$ state's concurrence is equivalent to the Bell $\ket{\psi^{+}}$ and $\ket{\psi^{-}}$ states under non-Markovian AD noise without WM whereas the $\ket{NS_2}$ state is the most robust among all the considered states. Similarly, variations of the teleportation fidelity~\cite{horodecki1996teleportation} and fidelity deviation~\cite{ghosal2020optimal} for all the considered states under non-Markovian AD channel are provided in Fig.~\ref{TFid_NMAD} and Fig.~\ref{FD_NMAD} respectively. Out of all the Bell states and two-qubit negative quantum states, the $\ket{NS_2}$ state is the most suitable state for universal quantum teleportation with WM under both non-Markovian AD and RTN channels. 
\end{document}